\documentclass[twocolumn,floatfix,pra,nofootinbib]{revtex4-1}
\usepackage[colorlinks=true,citecolor=blue]{hyperref}
\usepackage{graphicx}
\usepackage{yquant,braket}
\usetikzlibrary{fit, quotes}
\usepackage{physics}
\usepackage{amsmath}
\usepackage{amssymb}
\usepackage{amsthm}
\usepackage{stackrel}
\usepackage{comment}
\usepackage{todonotes}
\usepackage{color}
\usepackage{dsfont}
\usepackage{mathrsfs}
\usepackage{bm}
\usepackage{units}
\usepackage{algorithm} 
\usepackage[noend]{algpseudocode}
\usepackage{float}
\usepackage[margin=0pt]{subfig}
\usepackage{upgreek}
\usepackage[shortlabels]{enumitem}
\usepackage{bbold}
\usepackage[normalem]{ulem} 
\usepackage{academicons}
\usepackage[capitalise]{cleveref}
\crefformat{section}{\S#2#1#3}

\usepackage{enumitem}
\newlist{todolist}{itemize}{2}
\setlist[todolist]{label=$\square$}
\usepackage{pifont}

\definecolor{orcidlogocol}{HTML}{A6CE39}
\usepackage{xcolor}

\DeclareRobustCommand\rvdots{%
\vbox{%
\baselineskip4\p@\lineskiplimit\z@%
\kern-\p@%
\hbox{.}\hbox{.}\hbox{.}%
}%
}

\begin{document}

\title{Noise tailoring for Robust Amplitude Estimation}
\author{Archismita Dalal}
\email{archismita.dalal@zapatacomputing.com}
\author{Amara Katabarwa}
\email{amara@zapatacomputing.com}
\affiliation{Zapata Computing, Boston, MA 02110 USA}

\date{\today}

\begin{abstract}
A universal fault-tolerant quantum computer holds the promise to speed up computational problems that are otherwise intractable on classical computers; however, for the next decade or so, our access is restricted to noisy intermediate-scale quantum (NISQ) computers and, perhaps, early fault tolerant (EFT) quantum computers.
This motivates the development of many near-term quantum algorithms including robust amplitude estimation (RAE), which is a quantum-enhanced algorithm for estimating expectation values. 
One obstacle to using RAE has been a paucity of ways of getting realistic error models incorporated into this algorithm. 
So far the impact of device noise on RAE is incorporated into one of its subroutines as an exponential decay model, which is unrealistic for NISQ devices and, maybe, for EFT devices; this 
hinders the performance of RAE.
Rather than 
trying to
explicitly 
model
realistic noise effects, which 
may be
infeasible, we circumvent this obstacle by tailoring device noise to generate an effective noise model, whose impact on RAE closely resembles that of the exponential decay model.
Using noisy simulations, we show that our noise-tailored RAE algorithm is able to regain improvements in both bias and precision that are expected for RAE.
Additionally, on IBM's quantum computer our algorithm demonstrates advantage over the standard estimation technique in reducing bias.
Thus, our work extends the feasibility of RAE on NISQ computers, consequently bringing us one step closer towards achieving quantum advantage using these devices.

\end{abstract}
\maketitle

\section{Introduction}
Amazing progress and development of quantum processors and the paradigm of hybrid quantum-classical algorithms have worked in tandem to spark the age of near-term quantum computing.
The two major events that acted as catalysts for this noisy intermediate-scale quantum (NISQ) era are the invention of variational quantum eigensolver (VQE)~\cite{peruzzo2014variational} and publicly-available access to IBM's quantum devices through their cloud platform~\cite{ibmresearch2016}.
In addition to VQE for quantum chemistry, with promising experimental demonstrations~\cite{googlehartreefockvqe,Kandala2017, OMalley2016a, Hempel2018a}, variational quantum algorithms have applications in diverse fields including discrete optimization \cite{Farhi2014c,Bravyi2020, Egger2021, Akshay2019}, quantum machine learning and generative modeling \cite{Schuld2020, Schuld2021,Dallaire-Demers2018a,Cao2021,Liu2018, Benedetti2019, Perdomo-Ortiz2018, Alcazar2020, Rudolph2020b} and quantum error correction \cite{Johnson2017a}. 

Extensive studies to elucidate the viability of quantum advantage in near-term quantum computing have produced mixed results so far~\cite{Rudolph2020b,Huang2021, Huang2021a,Alcazar2021,Somma2021,Jaderberg2021,Wecker2015b}. 
A recent analysis argued that VQE will fail to provide quantum advantage over state-of-the-art quantum chemistry algorithms for estimating the ground-state energy of an industry-scale problem Hamiltonian~\cite{Gonthier2020a}.
This drawback of VQE arises because evaluating even a single energy within chemical accuracy requires a large number of samples, which was estimated to take up to days for some realistic molecules. 
Specifically, the number of samples required for a desired accuracy of~$\epsilon$ scales as $\mathcal{O}(\frac{1}{\epsilon^2})$, leading to huge run-times for VQE.
This lesser-acknowledged problem of infeasible runtimes, which was dubbed the ``measurement problem"~\cite{Gonthier2020a}, poses a major roadblock for hybrid algorithms, as well as many non-variational algorithms, towards achieving quantum advantage on devices in the NISQ era and beyond. 

In an effort to deal with the measurement problem on NISQ devices, recent works apply techniques of quantum amplification and quantum estimation borrowed from the field of fault-tolerant quantum computing~\cite{Wang2019a,Wang2021,Giurgica-Tiron2020,Tanaka2021}.
As a first step towards this effort, Ref.~\cite{Wang2019a} achieves a reduction in the desired number of samples for VQE by improving the sample scaling from $\mathcal{O}(\frac{1}{\epsilon^2})$ to $\mathcal{O}( \frac{2}{1-\alpha}(\frac{1}{\epsilon^{2(1-\alpha)}}-1))$, where $\alpha \in (0, 1]$.
More generally, these algorithms employ short-depth versions of quantum amplitude estimation~\cite{Brassard2002, Abrams1999a} to improve scaling and reduce runtimes for VQE, thus yielding interpolating scaling between the standard and the quantum phase estimation sampling rates of $\mathcal{O}(\frac{1}{\epsilon^2})$ and $\mathcal{O}(\frac{1}{\epsilon})$, respectively.

Our work focuses on the near-term implementation of the quantum estimation algorithm~\cite{Wang2021}.
An essential component of this algorithm is the likelihood function, which is modelled as an exponentially-decaying periodic function with a decay parameter~$f$, over measurement data.
This ``fidelity" parameter depends on the two-qubit gate fidelity~($f_{2Q}$), the number of two-qubit gates~($D$) and the number of qubits~($n$), 
thus incorporating noise effects into algorithm design and making this algorithm robust.
A detailed analysis based on this model derives the total run-time
\begin{align}
t =& e \frac{f_{2Q}^{nD/2}}{2\Bar{p}^2 }\Bigg[\frac{nD \ln(1/f_{2Q})}{2\epsilon^2} + \frac{1}{\sqrt{3}\epsilon} \nonumber \\
&+ \sqrt{\left(\frac{nD \ln(1/f_{2Q})}{2\epsilon^2}\right)^2 + \left(\frac{2\sqrt{2}}{\epsilon}\right)^2} \Bigg],
\end{align}
where $\Bar{p}$ incorporates state preparation and measurement errors.
This expression highlights the interplay between fidelity (noise) and depth (quantum coherence) towards reducing runtimes and improving scaling of measurements with respect to accuracy.

\begin{figure}
    \centering
    \includegraphics[width=90mm, height=50mm]{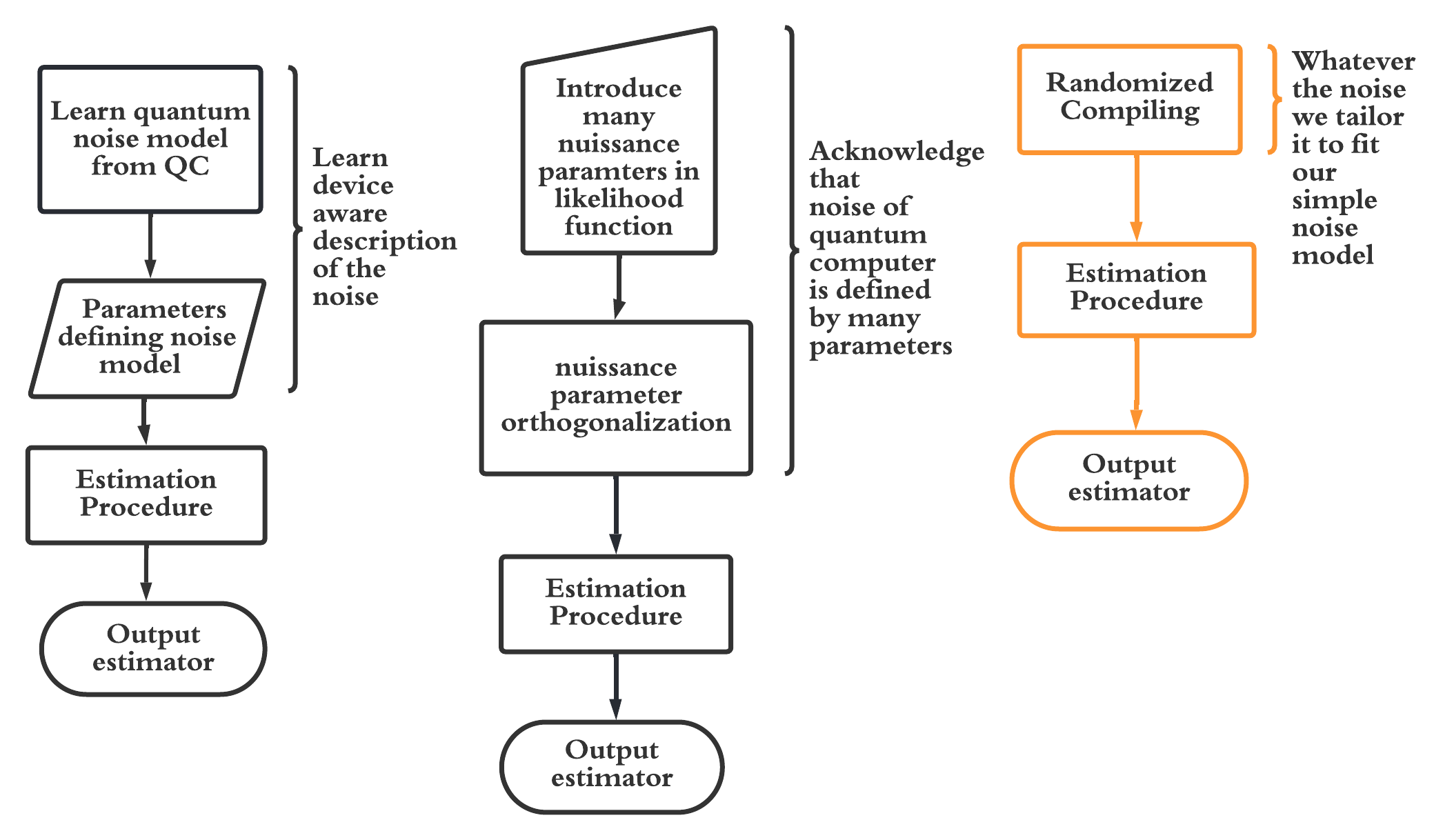}
    \caption{Three techniques for dealing with device noise in quantum amplitude estimation: (a) construct accurate noise models~\cite{Herbert2021},
    (b) add more parameters to the existing noise model~\cite{Tanaka2022},
    (c) tailor device noise into existing noise model~[{\color{blue}this work}].}
    \label{fig:three_philosophies}
\end{figure}

In practice, using quantum coherence to improve sampling efficiency requires running deeper quantum circuits, but their efficiency might be curbed by noise~\cite{katabarwa2021reducing,Rao2020,Giurgica-Tiron2021, Herbert2021}.
Additionally, more complex errors originating from unpredictable sources, including miscalibration of gates and undesired correlations between qubits, in NISQ devices can deviate the likelihood function from the assumed exponential decay model.
To this end, there have been two techniques so far for improving the performance of amplitude estimation in such noisy scenarios; see \cref{fig:three_philosophies}.
The first one is to experimentally approximate the decay parameter and insert it into the likelihood function~\cite{Herbert2021}, and the second is to introduce more decay parameters into the likelihood function~\cite{Tanaka2022}. 
In summary, one either measures the nuisance parameters in a simplified noise model or adds extra nuisance parameters in the likelihood function.

Our approach is to tailor the device noise into stochastic noise, thus producing an effective noise model closer to the exponential decay.
While~\cref{fig:three_philosophies} highlights three different methods to make quantum amplitude estimation feasible for near-term applications, we emphasize that they are not mutually exclusive and some combined procedure might eventually be the best.
Using noisy simulations, we show that the measurement data from noise-tailored amplitude amplification circuits come closer to the exponential decay model for likelihood function. 
Furthermore, we assess the performance of our technique at the task of estimating expectation values using current quantum processors.

Our paper is organized as follows. 
We begin in \S\ref{sec:rae} by briefly reviewing the robust amplitude estimation algorithm, and then inspect the impact of both incoherent and coherent errors on the likelihood function.
In \S\ref{sec:rc} we summarize randomized compiling, an efficient noise-tailoring technique for quantum devices, and study its effect on this likelihood function.
Finally we present our results in~\S\ref{sec:rc-rae} on the success and limitations of this noise-tailoring technique for robust amplitude estimation using IBM devices, both simulators and real hardware, and conclude our work in \S\ref{sec:conclusion}.


\section{Noise in Robust Amplitude Estimation}
\label{sec:rae}
Robust amplitude estimation (RAE) is a quantum-enhanced algorithm for estimating the expectation value of a Hermitian operator with higher accuracy and precision as compared to the standard sampling technique of direct averaging~\cite{katabarwa2021reducing}.
In this section, we first briefly explain the RAE algorithm, then we introduce mathematical models describing major sources of incoherent and coherent errors in superconducting qubits.
Finally, we discuss the impact of these errors on the RAE algorithm.
\begin{figure}
\centering
    \subfloat[An enhanced sampling circuit with three Grover iterates ($G$).]{
\begin{tikzpicture}
\begin{yquant}
qubit {$\ket0^{\otimes n}$} q;
box {$A$} q[0];
box {\hbox to .6cm{\hfil$G$\hfil}} q[0];
box {\hbox to .6cm{\hfil$G$\hfil}} q[0];
box {\hbox to .6cm{\hfil$G$\hfil}} q[0];
dmeter {P} q[0];
\end{yquant}
\end{tikzpicture}
} \\
    \subfloat[Circuit for $G$.]{
    \begin{tikzpicture}
\begin{yquant}
qubit {} q;
box {$P$} q[0];
box {$A^\dagger$} q[0];
box {$R_0$} q[0];
box {$A$} q[0];
\end{yquant}
\end{tikzpicture}
    }
\caption{The structure and components of a n-qubit quantum circuit used in estimating $\langle A| P |A \rangle$ by enhanced sampling. $A$ is a parametrized quantum circuit, $P$ is the operator to be measured and $R_0$ creates reflection about $\ket0^{\otimes n}$.}
\label{fig:es-circuit}
\end{figure}
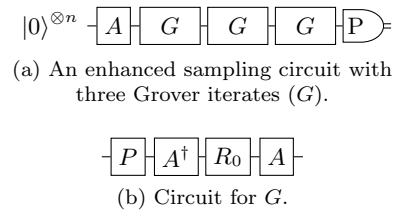

\subsection*{RAE algorithm}
Given a $n$-qubit Hermitian operator~$P$ with eigenvalues $\pm 1$ and a quantum state~$\ket{A} = A \ket{0}^{\otimes n}$, the goal is to estimate the expectation value~$\Pi := \langle A| P |A \rangle$.
To do this, the RAE algorithm uses measurement data from enhanced sampling circuits~\cite{Wang2021}, and then performs the classical postprocessing of maximum likelihood estimation (MLE)~\cite{katabarwa2021reducing}.
For a noisy enhanced sampling circuit with $L$ layers, each layer resembles a Grover iterate for quantum amplification and has a fidelity~$f \in [0,1]$; see \cref{fig:es-circuit}.
Furthermore, the likelihood of obtaining a measurement bit-string with parity $d \in \{0 ,1\}$ is
\begin{equation}
\label{eq:noisyCheby}
    \mathcal{L}(d|\Pi)=\frac{1}{2}\left(1 +  (-1)^d f^{(L + \nicefrac{1}{2})}\operatorname{cos}((2L+1)\operatorname{cos}^{-1}(\Pi))\right).
\end{equation}
The layer fidelity parameter is in-turn related to the nuisance parameter~$\lambda$ of the exponential decay noise model as $f=e^{-\lambda}$, thus incorporating the impact of noise into the above likelihood function.
Combining likelihood functions for enhanced sampling circuits with different $L$s and subsequently using MLE technique, the RAE algorithm yields an estimate of $\Pi$.

It was empirically shown that RAE affords not only a way to improve the precision of one's estimate but also a way to mitigate the bias in one's estimate at least for a two-qubit system~\cite{katabarwa2021reducing}.
The crucial insight behind this observation is that as long as the noise model describing the quantum hardware is not far from the assumed noise model for RAE, i.e. the exponential decay model, or has the same effect on the algorithm as this assumed noise model, the effect of the noise in MLE is simply to slow the rate of information gain as opposed to biasing the estimates.
In this work, we push this insight further and endeavour to tailor the real and complicated noise on a quantum hardware to produce an effective noisy circuit so that $f$ in likelihood function (\ref{eq:noisyCheby}) becomes a better approximation to the effect of noise in the estimation process.

\subsection*{Relevant noise models}
For our simulations, we use the model of a noisy superconducting quantum processor and include incoherent errors from amplitude damping and dephasing and coherent errors from residual $ZZ$ interactions between qubits, which are the major sources of error for these processors.
Although amplitude and phase damping errors are bad, coherent error is the major performance-limiting factor for quantum algorithms because these errors can accumulate adversarially with increasing circuit depth.
Ref.~\cite{Karamlou2021} demonstrates that, for superconducting quantum processors, the residual $ZZ$ errors exert a detrimental impact on hybrid algorithms using parameterized quantum circuits (PQC).

The combined effect of phase and amplitude damping errors is described by a stochastic error channel.
For a single-qubit, with decay rate~$T_1$, dephasing rate~$T_2$, and a time-step parameter~$t_\text{step}$, this channel is defined using the three Kraus matrices
\begin{align}
        E_1 &= \begin{pmatrix}
             1 & 0 \\
             0 & \sqrt{1 - \alpha -\beta}
         \end{pmatrix}, \nonumber \\
        E_2 &= \begin{pmatrix}
              0 & \sqrt{\alpha} \\
              0 & 0
            \end{pmatrix}, \nonumber \\
        E_3 &= \begin{pmatrix}
              0 & 0 \\
              0 & \sqrt{\beta}
            \end{pmatrix},
  \end{align}
where $\alpha = 1 - e^{-\frac{t_\text{step}}{T_1}}$ and $\beta = 1-e^{-\frac{ t_\text{step}}{T_2}}$.
Extending this formulation to $n$ qubits, the incoherent error channel transforms a $n$-qubit density matrix $\rho$ into 
 \begin{equation}
 \label{eq:incoherent_noise}
      \Lambda (\rho) = \sum_{i_1=1}^{i_1=3} 
      \cdots\sum^{i_n=3}_{i_n=1} E_{i_1} 
      \otimes 
      \dots  \otimes E_{i_n}  \rho  E_{i_1}^{\dagger} 
      \otimes 
      \dots  \otimes  E_{i_n}^{\dagger}.
  \end{equation} 
For our simulations, we use the implementation of this error channel provided by IBM's \texttt{qiskit}~\cite{T1T2noise}.
  
The coherent error arising from undesired $ZZ$ interactions between transmon qubits is modelled as a modification to the ideal CNOT gate~\cite{Karamlou2021}.
These $ZZ$ interactions are produced by small anharmonicities in the qubits.
The overall effect of these interactions on a pair of qubits, namely control~$c$ and target~$t$, is expressed as
\begin{align}
  \Xi = \xi_{c, t} (ZZ)_{c,t} + \sum_{i}^{|\text{spec}(c)|} \xi_{c,i} (ZZ)_{c,i}
  + \sum_{j}^{|\text{spec}(t)|} \xi_{t,j} (ZZ)_{t, j},
  \end{align} 
where $\text{spec}(c)$ and $\text{spec}(t)$ are the sets of qubits coupled to $c$ and $t$, respectively, with $\xi_{c,i}$ and $\xi_{t,j}$ being the corresponding $ZZ$ interactions. 
In the presence of these coherent errors, the expression for CNOT over gate time~$t_\text{gate}$, in terms of the native cross-resonance gate~\cite{ChadRigetti2010, JerryChow2011}, is modified as
\begin{align}
  \label{eq:coherent_noise}
     \text{CNOT} &:= e^{-\frac{i \pi}{4}} e^{i\frac{\pi}{4}Z \otimes I} e^{-i\frac{\pi}{4} Z \otimes X   } e^{i\frac{\pi}{4}I \otimes X} \nonumber\\
     &\longmapsto e^{-\frac{i \pi}{4}} e^{i\frac{\pi}{4}Z \otimes I} e^{i(\Gamma Z \otimes X  + \Xi)t_\text{gate} } e^{i\frac{\pi}{4}I \otimes X},
 \end{align}
where $\Gamma = \frac{\pi}{4t_\text{gate}}$ is the coupling between the control and target qubits.

\subsection*{Impact of noise on the likelihood function}
\begin{figure}
\begin{tikzpicture}
   \begin{yquant}
     qubit {$\ket{\reg_{\idx}}$} q[4];
     box {$X$} q[0];
     box {$X$} q[1];
     box {$R_y(\theta_0)$} q[2];
     cnot q[0]| q[2];
     cnot q[1] | q[2];
     cnot q[3] | q[2];
   \end{yquant}
\end{tikzpicture}
\caption{A 4-qubit parameterized circuit for Hydrogen molecule in a minimal basis.}
\label{fig:4_qubit_circuit}
\end{figure}
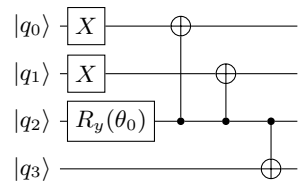
\begin{figure}[]
    \centering
    \subfloat[Fitting simulation results to $P(0|\Pi=0.2238,L)$ in the presence of only incoherent noise: amplitude damping with $T_1=84~\upmu$s and phase damping with $T_2=110~\upmu$s.]{
    \includegraphics[width=.85\linewidth]{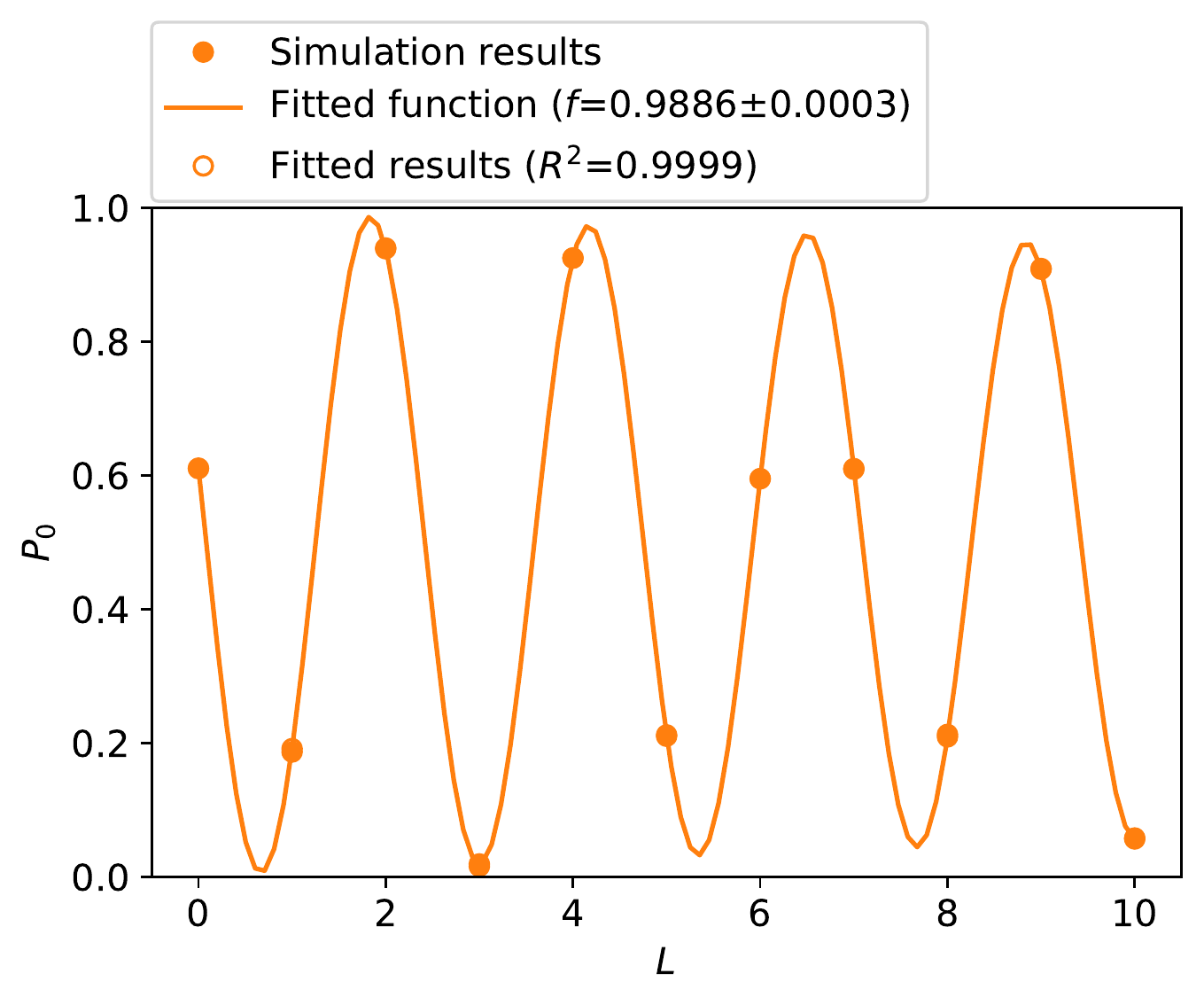}} \\
    \subfloat[Fitting simulation results to $P(0|\Pi=0.2238,L)$ in the presence of both incoherent and coherent noise: amplitude damping with $T_1=84~\upmu$s, phase damping with $T_2=110~\upmu$s and residual $ZZ$ interactions with~$\nicefrac{\xi}{2\pi}=45$~kHz.]{
    \includegraphics[width=.85\linewidth]{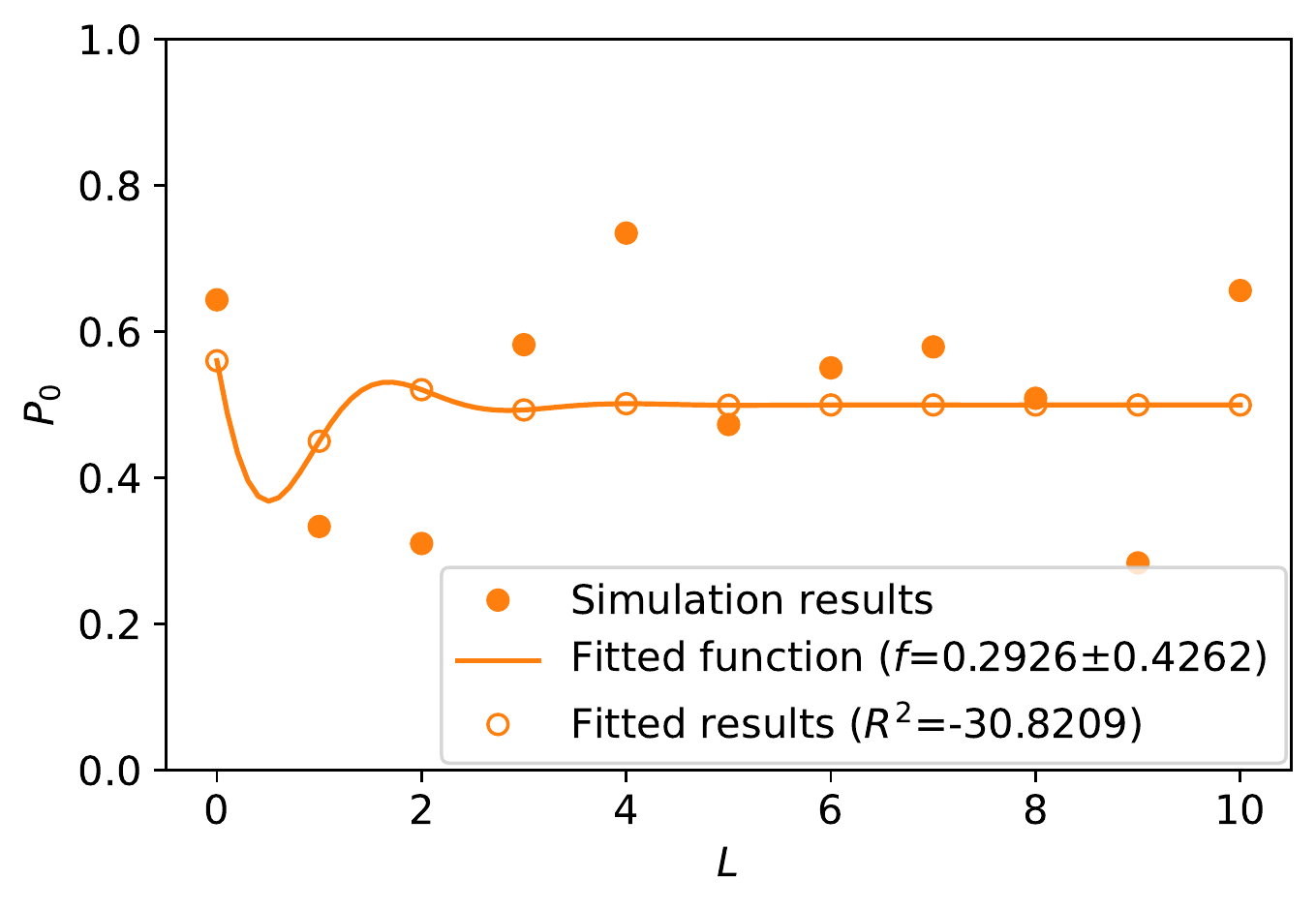}}
    \caption{Impact of incoherent and coherent noise on the likelihood function $P(0|\Pi,L)$~\eqref{eq:P_0} obtained from noisy simulations of enhanced sampling circuits, with $A$ being the 4-qubit PQC~(\cref{fig:4_qubit_circuit}) and $\Pi=\langle XXXX \rangle=0.2238$.}
    \label{fig:increaseL}
\end{figure}

To understand the effect of incoherent and coherent errors on the RAE algorithm, we investigate how these errors impact the likelihood function~\eqref{eq:noisyCheby}.
To do this, we use the likelihood of getting even parities as
\begin{equation}
\label{eq:P_0}
    P(0|\Pi,L)
    =\frac{1}{2}\left(1 + f^{(L + \nicefrac{1}{2})}\operatorname{cos}((2L+1)\operatorname{cos}^{-1}\Pi)\right),
\end{equation}
where $L=0$ is the standard sampling case.
We perform noisy simulations of enhanced sampling circuits with $A$ being the PQC describing a 4-qubit hydrogen molecule~(\cref{fig:4_qubit_circuit}), with $\theta_0=-6.057$, and $P$ being the 4-qubit Pauli operator~$XXXX$. 
In \cref{fig:increaseL}(a), we present results from simulations incorporating only amplitude and phase damping, with typical $T_1$ and $T_2$ values for IBM devices, for a fixed $\Pi$ and increasing $L$.
We observe that the likelihood values obtained from these simulations yield a good fit to the desired functional form~\eqref{eq:P_0}, as evident from the near-unity $R^2$ value.
This implies that despite the noise we should be able to get a good estimate of $f$ and the estimation process could be largely salvaged.

On the other hand, adding moderate level of coherent errors from $ZZ$ interactions drives the simulation results far away from our assumed functional form with little hope of getting a good estimate of $f$; see \cref{fig:increaseL}(b).
This observation serves as a strong evidence of how coherent errors impact likelihood values, and can consequently deteriorate the performance of RAE.
From these results we infer that with the assumed noise model behind the derivation of the likelihood function~\eqref{eq:noisyCheby}, RAE will not always offer the sought after sampling power.

\section{Randomized Compiling for RAE}
\label{sec:rc}
Motivated by our findings on how detrimental coherent errors are for likelihood values, we aim to engineer an effective noise profile whose impact on RAE is similar to that of an exponential decay model. 
To this end, we employ randomized compiling (RC) that efficiently tailors coherent errors into stochastic errors~\cite{Wallman2016}.
In this section, we first briefly state the background of RC, followed by an explanation of the basic procedure for RC.
Finally, we study the impact of this technique on likelihood values in the presence of different magnitudes of coherent error.

The idea of RC was originally proposed in the context of quantum error correction and fault tolerant quantum computing, where thresholds for different error-correcting codes are estimated with an assumption of stochastic Pauli noise~\cite{Aliferis2006,Aliferis2008}.
These estimates can be way off if coherent errors are present \cite{AndrewDarmawan2017} since they  pile up systematically.
In this regard, RC was introduced to construct an effective quantum channel, which is described by only stochastic incoherent noise.
Recently, RC has been used for improving algorithmic performances of NISQ devices~\cite{Ville2021, Hashim2021,Urbanek2021,Ferracin2022,GMFL22}.
Thus, besides making RAE feasible for NISQ devices, our work adds on to the growing field of research in noise mitigation and noise tailoring within the NISQ era.

\subsection*{RC procedure}
\begin{figure}[]
\subfloat[A 2-qubit $\mathcal{C}_\text{bare}$ with alternating layers of 3 easy cycles ($\Vec{\mathcal{E}_1}$, $\Vec{\mathcal{E}_2}$ and $\Vec{\mathcal{E}_3}$) and 2 hard cycles ($\mathcal{H}_1$ and $\mathcal{H}_2$).]{
\begin{tikzpicture}
\begin{yquant}
qubit {$\ket{\reg_{\idx}}$} q[2];
[fill=red!20]
box {${\mathcal{E}_{11}}$} q[0];
[fill=red!20]
box {${\mathcal{E}_{12}}$} q[1];
[fill=red!20]
box {$\mathcal{H}_1$} (q[0], q[1]);
[fill=red!20]
box {${\mathcal{E}_{21}}$} q[0];
[fill=red!20]
box {${\mathcal{E}_{22}}$} q[1];
[fill=orange!20]
[fill=red!20]
box {$\mathcal{H}_2$} (q[0], q[1]);
[fill=red!20]
box {${\mathcal{E}_{31}}$} q[0];
[fill=red!20]
box {${\mathcal{E}_{32}}$} q[1];
\end{yquant}
\end{tikzpicture}
}
\\
\subfloat[A random duplicate with $\Vec{P}_i$ sampled from $\mathcal{P}_2$ and $\Vec{P}_{i}^c = \mathcal{H}_{i}\Vec{P}_{i}^{\dagger} \mathcal{H}_{i}^{\dagger}$.
We assume that $\mathcal{P}_0^c = \mathbb{I}^{\otimes 2}$.]{
\begin{tikzpicture}
    \begin{yquant}
    qubit {$\ket{\reg_{\idx}}$} q[2];
    [fill=red!20]
    box {$\mathcal{E}_{11}$} q[0];
    [fill=red!20]
    box {$\mathcal{E}_{12}$} q[1];
    [fill=orange!20]
    box {$P_{11}$} q[0];
    [fill=orange!20]
    box {$P_{12}$} q[1];
    [fill=red!20]
    box {$\mathcal{H}_1$} (q[0], q[1]);
    [fill=orange!20]
    box {$P_{11}^{c}$} q[0];
    [fill=orange!20]
    box {$P_{12}^{c}$} q[1];
    [fill=red!20]
    box {$\mathcal{E}_{21}$} q[0];
    [fill=red!20]
    box {$\mathcal{E}_{22}$} q[1];
    [fill=orange!20]
    box {$P_{21}$} q[0];
    [fill=orange!20]
    box {$P_{22}$} q[1];
    [fill=red!20]
    box {$\mathcal{H}_2$} (q[0], q[1]);
    [fill=orange!20]
    box {$P_{21}^{c}$} q[0];
    [fill=orange!20]
    box {$P_{22}^{c}$} q[1];
    [fill=red!20]
    box {$\mathcal{E}_{31}$} q[0];
    [fill=red!20]
    box {$\mathcal{E}_{32}$} q[1];
    \end{yquant}
\end{tikzpicture} 
} \\
\subfloat[The circuit is re-compiled to preserve the depth of $\mathcal{C}_\text{bare}$. The new easy cycles $\Vec{\widetilde{\mathcal{E}_i}} = \Vec{P}_i \Vec{\mathcal{E}_i} \Vec{P}_{i-1}^c $]{
\begin{tikzpicture}
\begin{yquant}
qubit {$\ket{\reg_{\idx}}$} q[2];
[fill=red!20]
box {$\widetilde{\mathcal{E}_{11}}$} q[0];
[fill=red!20]
box {$\widetilde{\mathcal{E}_{12}}$} q[1];
[fill=red!20]
box {$\mathcal{H}_1$} (q[0], q[1]);
[fill=red!20]
box {$\widetilde{\mathcal{E}_{21}}$} q[0];
[fill=red!20]
box {$\widetilde{\mathcal{E}_{22}}$} q[1];
[fill=orange!20]
[fill=red!20]
box {$\mathcal{H}_2$} (q[0], q[1]);
[fill=red!20]
box {$\widetilde{\mathcal{E}_{31}}$} q[0];
[fill=red!20]
box {$\widetilde{\mathcal{E}_{32}}$} q[1];
\end{yquant}
\end{tikzpicture}
}
\caption{Circuit representations for logically-equivalent circuits before, during and after RC.}
\label{fig:random_duplicate_circuit}
\end{figure}
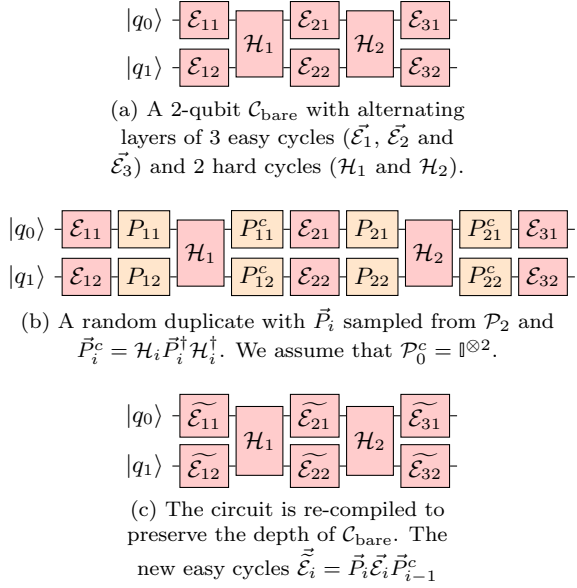

Given a $n$-qubit ``bare" circuit $\mathcal{C}_\text{bare}$, RC produces a set of $N$ random circuits $\{\tilde{\mathcal{C}_\ell}\}~\forall \ell \in \{1, 2, \cdots, N\}$ that are all logically equivalent to $\mathcal{C}_\text{bare}$.
Logical equivalence means that all these circuits produce exactly the same wave-function in the noiseless limit.
Each $\tilde{\mathcal{C}_\ell}$, which we refer to as a ``random duplicate", is constructed by the following procedure:
\begin{enumerate}
    \item Partition $\mathcal{C}_\text{bare}$ into ``easy" and ``hard" cycles, where we define an easy cycle to be a layer of single-qubit gates and a hard cycle to be a layer of two-qubit entangling gates,
    \item Uniformly sample from the $n$-qubit Pauli group~$\mathcal{P}_n$ to get an element,
    \item Insert the element  after an easy cycle,
    \item Insert gates after a hard cycle to ensure the logical equivalence to $\mathcal{C}_\text{bare}$.
\end{enumerate}
Additionally, we combine all adjacent single-qubit gates to preserve the circuit depth.
Using a convenient tensor product notation
\begin{equation}
    \Vec{\mathcal{E}_i} =\mathcal{E}_{i1} \otimes \mathcal{E}_{i2} \otimes \cdots \otimes \mathcal{E}_{in},
\end{equation}
where $\mathcal{E}_{ij}$ is the single-qubit gate acting on qubit~$j$ in an easy cycle~$i$, Fig.~\ref{fig:random_duplicate_circuit} shows the construction and structure of a random duplicate for a 2-qubit circuit.

In our implementation, the total number of measurements obtained by executing the RC procedure is kept equal to the number of measurements allocated to $\mathcal{C}_\text{bare}$.
By defining a tuple~$(\mathcal{C}, M)$ for a circuit~$\mathcal{C}$ with~$M$ measurements, we can thus formulate RC as the map
\begin{equation}
    f_\text{RC}:(\mathcal{C}_\text{bare},M) \longmapsto (\mathcal{C}_\text{RC},M) 
\end{equation}
where $(\mathcal{C}_\text{RC},M)$ represents a virtual circuit with measurements being the union of measurement outcomes from all random duplicates. 
For the set $\{(\tilde{\mathcal{C}_\ell}, M_\ell)\}$ and the identity $\sum_{\ell=1}^{N}M_\ell=M$, we choose
\begin{align}
    M_\ell &= \lfloor\nicefrac{M}{N}\rfloor,\; \forall \ell \in \{1, 2, \cdots, N-1\},\nonumber\\
    M_N &= M-\sum_{\ell=1}^{N-1}M_\ell.
\end{align}
All these random duplicates in the noisy setting are only distinguished by how the noise affects them since they are all equivalent in the noiseless setting. 
Then the intuition is that upon running them all on hardware and collecting the results we get bit-strings that behave as though they came from a circuit $\mathcal{C}_\text{RC}$ with no coherent noise but simply stochastic noise.

\subsection*{Impact of RC on the likelihood function}
\begin{figure}[]
    \centering
    \subfloat[Fitting simulation results (with and without RC) to $P(0|\Pi,L=1)$, for 1000 different parameterization of $A$, in the presence of both incoherent and coherent noise: amplitude damping with $T_1=84~\upmu$s, phase damping with $T_2=110~\upmu$s and residual $ZZ$ interactions with $\nicefrac{\xi}{2\pi}=75$kHz. ]
    {
    \includegraphics[width=.85\linewidth]{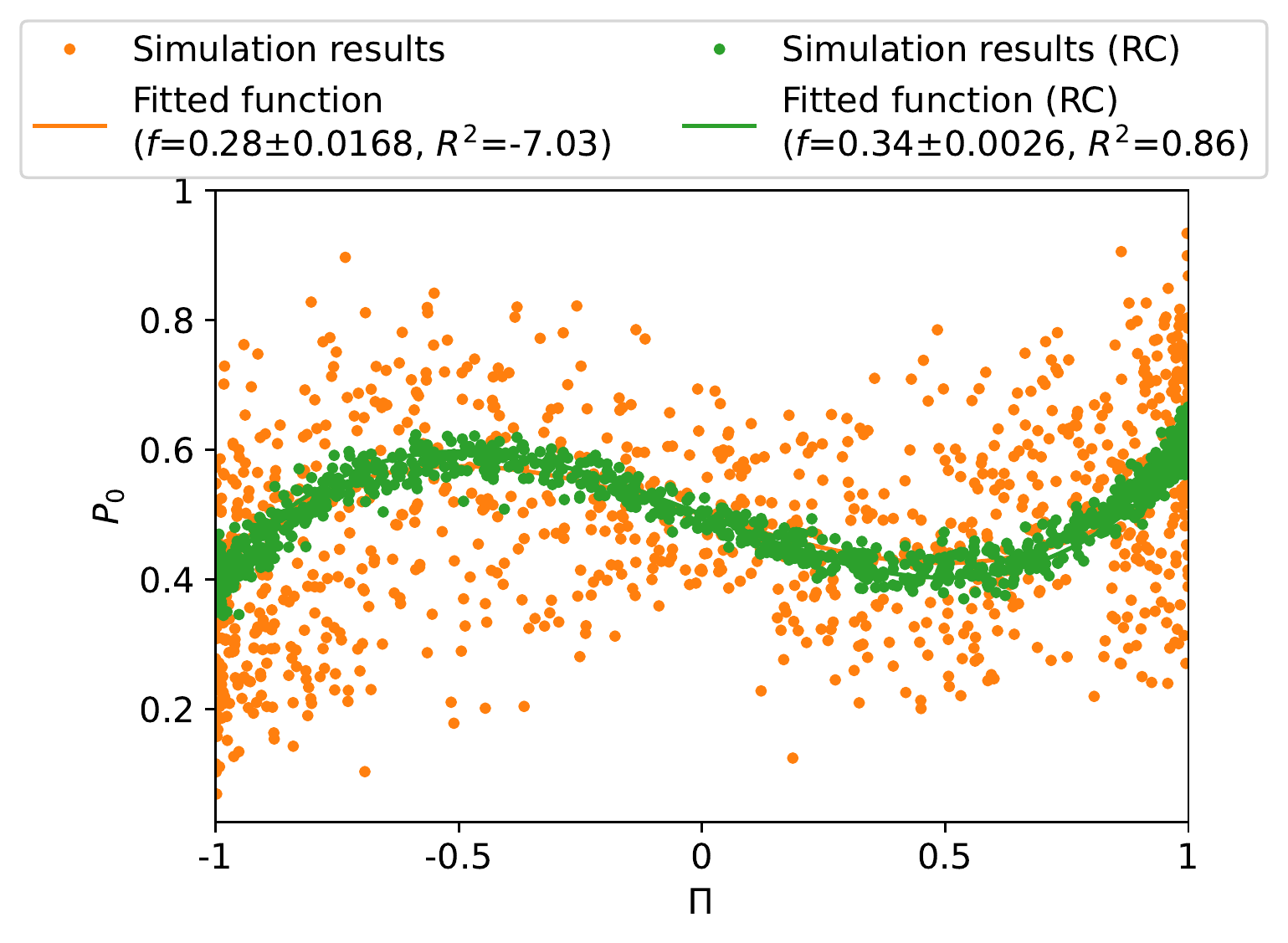}} \\
    \subfloat[Fitting simulation results (with and without RC) to $P(0|\Pi=0.2238,L)$ in the presence of both incoherent and coherent noise: amplitude damping with $T_1=84~\upmu$s, phase damping with $T_2=110~\upmu$s and residual $ZZ$ interactions with $\nicefrac{\xi}{2\pi}=10$kHz (circles) and~$\nicefrac{\xi}{2\pi}=75$~kHz (crosses).]{
    \includegraphics[width=.85\linewidth]{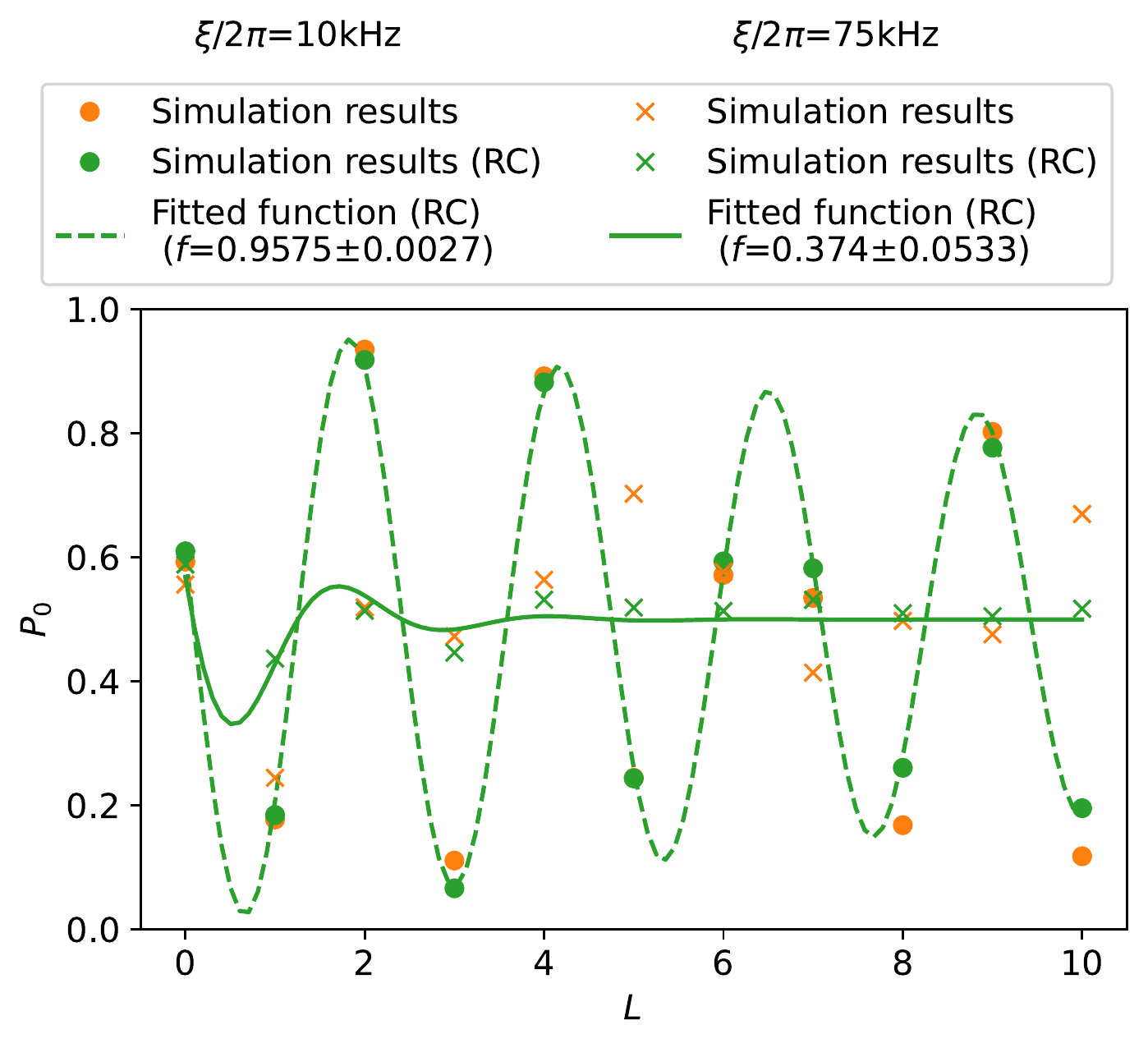}}
    \caption{Impact of the RC procedure on the likelihood function~$P(0|\Pi,L)$~\eqref{eq:P_0} obtained from noisy simulations of enhanced sampling circuits, with $A$ being the 4-qubit PQC~(\cref{fig:4_qubit_circuit}) and $P=XXXX$.
    }
    \label{fig:impact_of_rc}
\end{figure}
We study how RC impacts the likelihood of getting even parities~\eqref{eq:P_0} from noisy simulations of enhanced sampling circuits~(\cref{fig:es-circuit}), with the same $A$ and $P$ as in the previous section.
As this likelihood function depends on the exact expectation value~$\Pi$ and number of layers, i.e.\ Grover iterates,~$L$, our analysis is performed in the following two ways:
\begin{enumerate}
    \item Fix $L=1$ and vary $\Pi\in[-1,1]$ by choosing 1000 different values for $\theta_0$ in~\cref{fig:4_qubit_circuit},
    \item Fix $\Pi=0.2238$ and increase $L$ from 0 to 10.
\end{enumerate}
Furthermore, we generate $N=50$ random duplicates for each enhanced sampling circuit.

In~\cref{fig:impact_of_rc}(a), we observe that even in the presence of moderate $ZZ$ interactions with strength $\nicefrac{\xi}{2\pi}=75$~kHz, RC successfully tailors this coherent error into incoherent error for circuits with one Grover iterate, and consequently brings the likelihood function close to the desired form~\eqref{eq:P_0}.
The tailored circuits fit the function $P(0|\Pi,L=1)$ closely and yields a layer fidelity of 0.34, signifying that the resultant depolarizing channel is very noisy.
As we increase circuit depth beyond one Grover iterate, coherent noise piles up producing no signal, as seen in~\cref{fig:impact_of_rc}(b).
The resultant likelihood function is essentially flat and uninformative, with negligible enhanced sampling power; on the other hand for weak coherent noise, i.e.\ with $\nicefrac{\xi}{2\pi}=10$~kHz, RC is able to consistently tailor the likelihood values to the desired values for increasing $L$ and the resultant function~$P(0|\Pi=0.2238,L)$ is not very destructive.
From these observations we can hypothesize that RC can improve RAE in the presence of coherent noise as long as the noise is not very strong.

\section{RC-assisted RAE in practice}
\label{sec:rc-rae}
In this section, we present results on estimating Pauli expectation values using our technique of RC-assisted RAE.
This technique involves calculating maximum likelihood estimates using measurement data from randomly-compiled enhanced sampling circuits, as opposed to bare enhanced sampling circuits used in the RAE algorithm~\cite{katabarwa2021reducing}.
As RC tailors coherent error into stochastic error, consequently yielding a likelihood function resembling the desired functional form~\eqref{eq:noisyCheby}, we expect to obtain more precise and accurate estimates using the RC-assisted RAE algorithm.
We compare performances between three estimation techniques, namely RAE, RC-assisted RAE and direct averaging, for two different parameterized circuits.
In addition to the previously-used 4-qubit Hydrogen molecule~(\cref{fig:4_qubit_circuit}), we test these techniques for a 2-qubit low-depth circuit ansatz (LDCA)~\cite{Dallaire-Demers2019} with 20 CNOTs~(\cref{fig:2_qubit_circuit}).
Our tests are performed on both IBM simulator and a real device.

The classical postprocessing for MLE uses measurement data from multiple enhanced sampling circuits~(\cref{fig:es-circuit}) with increasing number of layers.
For a given maximum layer number~$L_\text{max}$, we calculate an estimate by collecting equal number of shots from circuits with $L\in\{0,1,\dots,L_\text{max}\}$ and then employing the MLE technique on this data.
By fixing the total number of shots to be $M$, we take $\lfloor\nicefrac{M}{L_\text{max}+1}\rfloor$ measurements from each circuit.
This is in contrast to techniques using $M$ measurements of each circuit~\cite{Giurgica-Tiron2021}, i.e. a total of $M(L_\text{max}+1)$ shots for each $L_\text{max}$, or optimizing number of shots for each circuit based on nuisance parameter~\cite{Herbert2021}.

For RC-assisted RAE, we also fix the number~$N$ of duplicates for increasing layers. 
An estimate~$\widetilde\Pi$ of the expectation value is taken as the average of estimates from $B$ Bayesian runs or MLEs, where the value of $B$ is judiciously chosen such that $\widetilde\Pi$ is converged.
Using 50 different measurement datasets, we obtain 50 estimates and then report their mean and root-mean-square error~(RMSE), along with their corresponding standard errors.
The total time, in units of $A$, required for calculating an estimate is the runtime
\begin{align}
    R =& \frac{M}{L_\text{max}+1}  \sum_{k=0}^{L_\text{max}}\left(\left(2k+1\right) +\frac{n_O}{n_A}k\right) \nonumber \\
    &= M\left(L_\text{max}+1\right)
    + \frac{Mn_O}{2n_A}L_\text{max},
\end{align}
where $n_O$ and $n_A$ are the number of two-qubit gates in $R_0$ and $A$, respectively.

\begin{widetext}

\begin{figure}[h!]
\begin{tikzpicture}
   \begin{yquant}
     qubit {$\ket{\reg_{\idx}}$} q[2];
     box {$X$} q[0];
     box {$R_z(\theta_0)$} q[0];
     box {$R_x(\frac{\pi}{2})$} q[0];
     box {$R_z(\theta_1)$} q[1];
     box {$H$} q[1];
     cnot q[1]| q[0];
     box {$R_z(2\theta_2)$} q[1];
     cnot q[1]| q[0];
     box {$R_x(-\frac{\pi}{2})$} q[0];
     box {$H$} q[0];
     box {$H$} q[1];
     box {$R_x(\frac{\pi}{2})$} q[1];
     cnot q[1]| q[0];
     box {$R_z(-2\theta_2)$} q[1];
     cnot q[1]| q[0];
     box {$H$} q[0];
     box {$R_x(-\frac{\pi}{2})$} q[1];
     cnot q[1]| q[0];
     box {$R_z(2\theta_3)$} q[1];
     cnot q[1]| q[0];
     box {$R_x(\frac{\pi}{2})$} q[0];
     box {$R_x(\frac{\pi}{2})$} q[1];
     cnot q[1]| q[0];
   \end{yquant}
 \end{tikzpicture}
 
 \vspace{2mm}
 \begin{tikzpicture}
   \begin{yquant}
    qubit {} q[2];
    box {$R_z(2\theta_4)$} q[1];
     cnot q[1]| q[0];
     box {$R_x(-\frac{\pi}{2})$} q[0];
     box {$H$} q[0];
     box {$R_x(-\frac{\pi}{2})$} q[1];
     box {$H$} q[1];
     cnot q[1]| q[0];
     box {$R_z(2\theta_4)$} q[1];
     cnot q[1]| q[0];
     box {$H$} q[0];
     box {$H$} q[1];
     box {$R_x(\frac{\pi}{2})$} q[0];
     box {$H$} q[1];
     cnot q[1]| q[0];
     box {$R_z(2\theta_5)$} q[1];
     cnot q[1]| q[0];
     box {$R_x(-\frac{\pi}{2})$} q[0];
     box {$H$} q[0];
     box {$H$} q[1];
     box {$R_x(\frac{\pi}{2})$} q[1];
     cnot q[1]| q[0];
     box {$R_z(-2\theta_5)$} q[1];
     cnot q[1]| q[0];
     box {$H$} q[0];
     box {$R_x(-\frac{\pi}{2})$} q[1];
   \end{yquant}
\end{tikzpicture}

\vspace{2mm}
\begin{tikzpicture}
   \begin{yquant}
     qubit {} q[2];
     cnot q[1]| q[0];
     box {$R_z(2\theta_6)$} q[1];
     cnot q[1]| q[0];
     box {$R_x(\frac{\pi}{2})$} q[0];
     box {$R_x(\frac{\pi}{2})$} q[1];
     cnot q[1]| q[0];
     box {$R_z(2\theta_7)$} q[1];
     cnot q[1]| q[0];
     box {$R_x(-\frac{\pi}{2})$} q[0];
     box {$H$} q[0];
     box {$R_x(-\frac{\pi}{2})$} q[1];
     box {$H$} q[1];
     cnot q[1]| q[0];
     box {$R_z(2\theta_7)$} q[1];
     cnot q[1]| q[0];
     box {$H$} q[0];
     box {$H$} q[1];
   \end{yquant}
 \end{tikzpicture}
\caption{A 2-qubit LDCA with
$\bm{\theta}=[ -1.491,  1.838,  1.977,  2.305, -3.124, 2.049, 1.254, -1.791]$.}
\label{fig:2_qubit_circuit}
\end{figure}
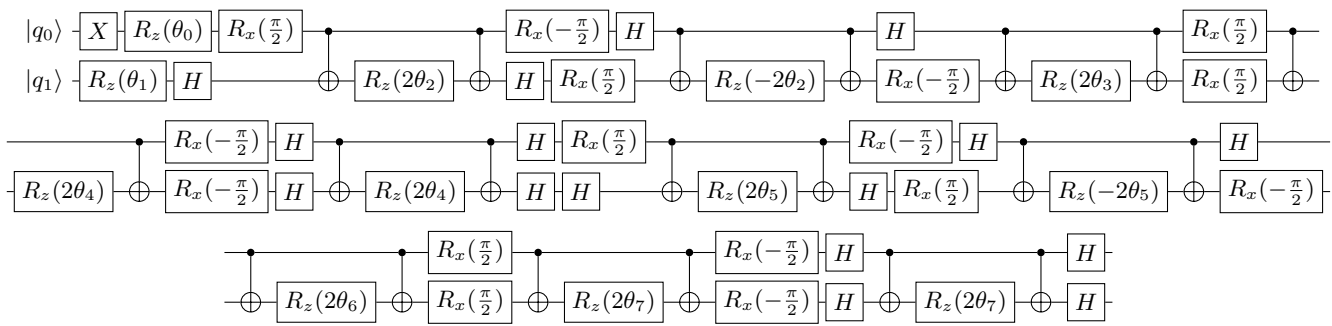
\end{widetext}


\subsection*{Performance on a noisy simulator}
\begin{figure}[]
    \begin{center}
    \subfloat[Estimation results in the presence of amplitude damping with $T_1=84~\upmu$s, phase damping with $T_2=110~\upmu$s and residual $ZZ$ interactions with~$\nicefrac{\xi}{2\pi}=10$~kHz.]{
    \includegraphics[width=.85\linewidth]{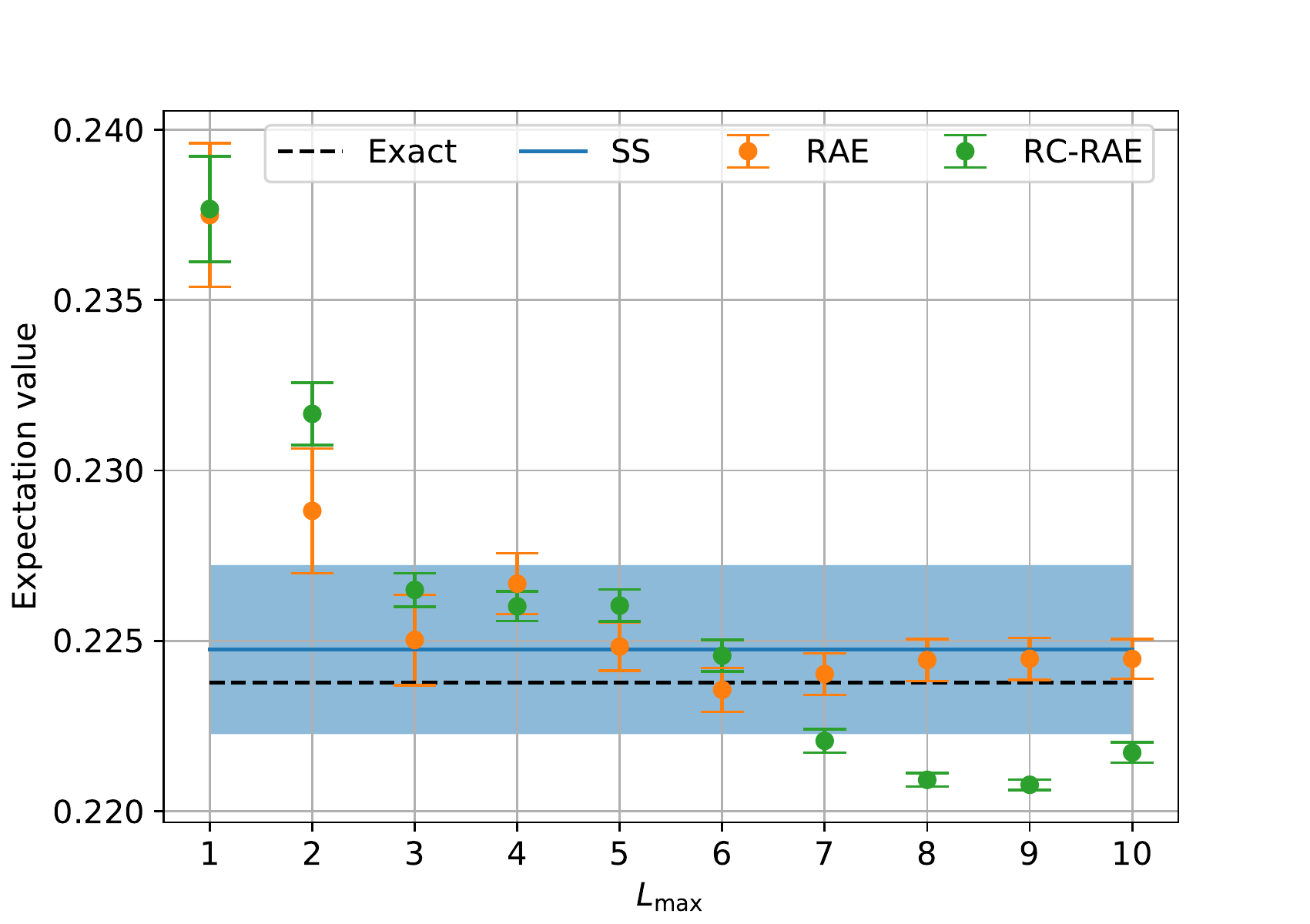}} \\
    \subfloat[Estimation results in the presence of amplitude damping with $T_1=84~\upmu$s, phase damping with $T_2=110~\upmu$s and residual $ZZ$ interactions with~$\nicefrac{\xi}{2\pi}=75$~kHz.]{
    \includegraphics[width=.85\linewidth]{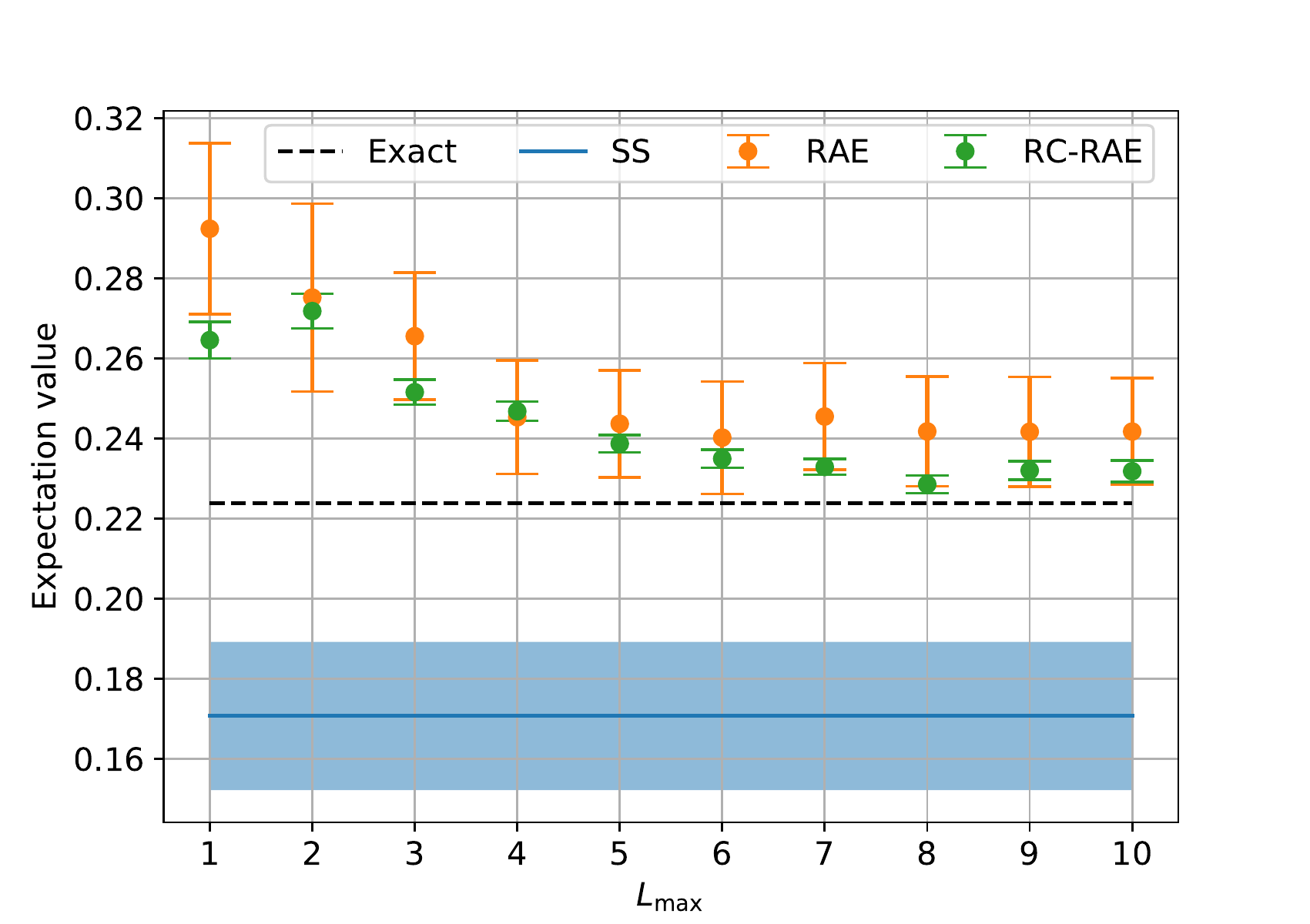}}  
    \end{center}
    \caption{Error mitigation properties of RAE and RC-assisted RAE (RC-RAE) as observed by simulating enhanced sampling circuits in the presence of both incoherent and coherent noise. Each point and its error bar represent the mean and standard error of 50 estimates of $ \Pi = \langle A|XXXX|A \rangle = 0.2238$, respectively, with $A$ being the 4-qubit PQC~(\cref{fig:4_qubit_circuit}).
    Each estimate is obtained from $B=500000$ runs and $M=20000$ shots, with RC using $N=50$.
    Blue line indicates the estimate obtained by direct averaging over $M$ shots from the SS circuit, and the blue band denotes the corresponding standard error.
  }
    \label{fig:compare_mitigate}
\end{figure}

\begin{figure}[]
    \centering
    \subfloat[Estimation results in the presence of amplitude damping with $T_1=84~\upmu$s, phase damping with $T_2=110~\upmu$s and residual $ZZ$ interactions with~$\nicefrac{\xi}{2\pi}=10$~kHz.]{
    \includegraphics[width=.85\linewidth]{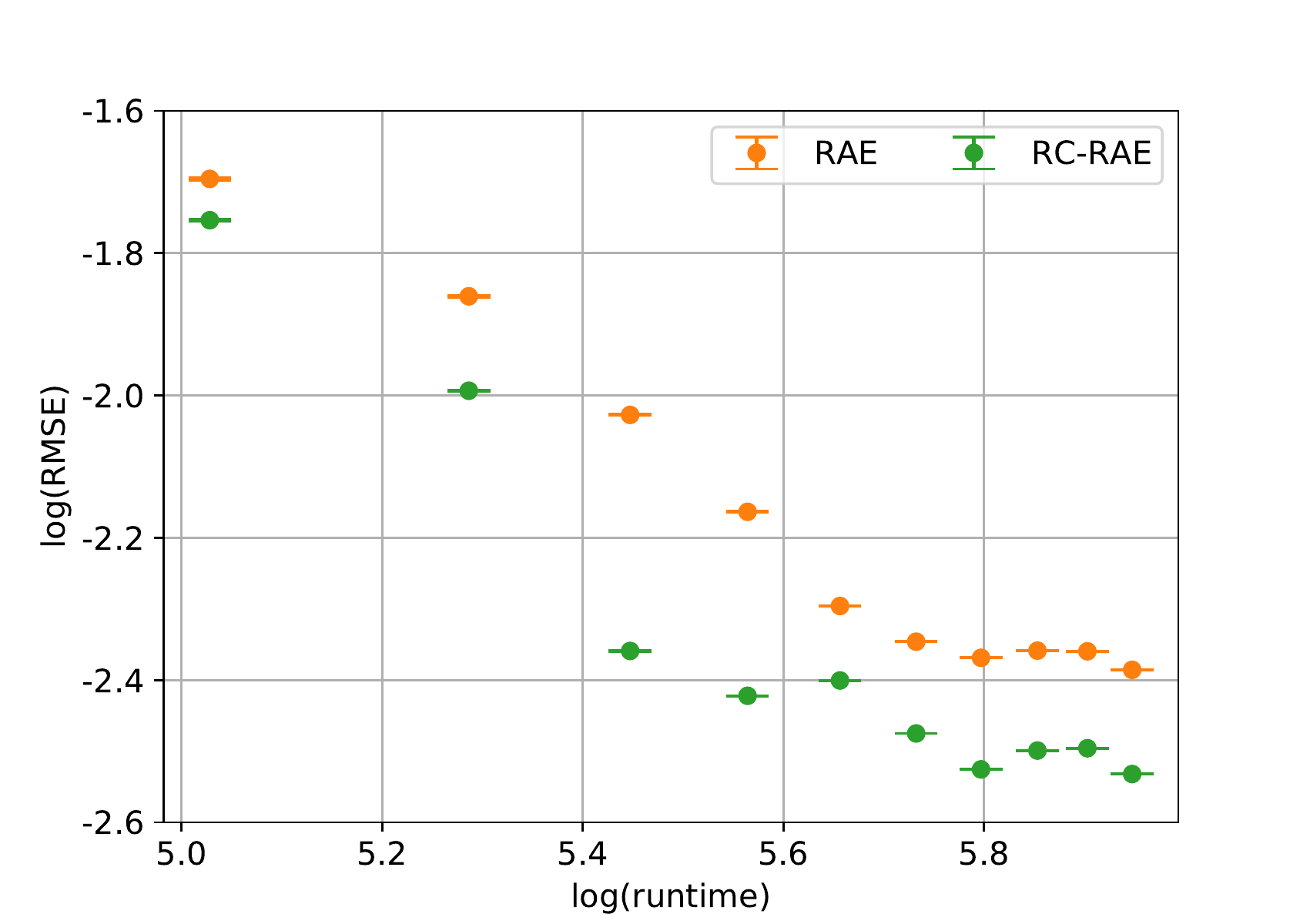}} \\
    \subfloat[Estimation results in the presence of amplitude damping with $T_1=84~\upmu$s, phase damping with $T_2=110~\upmu$s and residual $ZZ$ interactions with~$\nicefrac{\xi}{2\pi}=75$~kHz.]{
    \includegraphics[width=.85\linewidth]{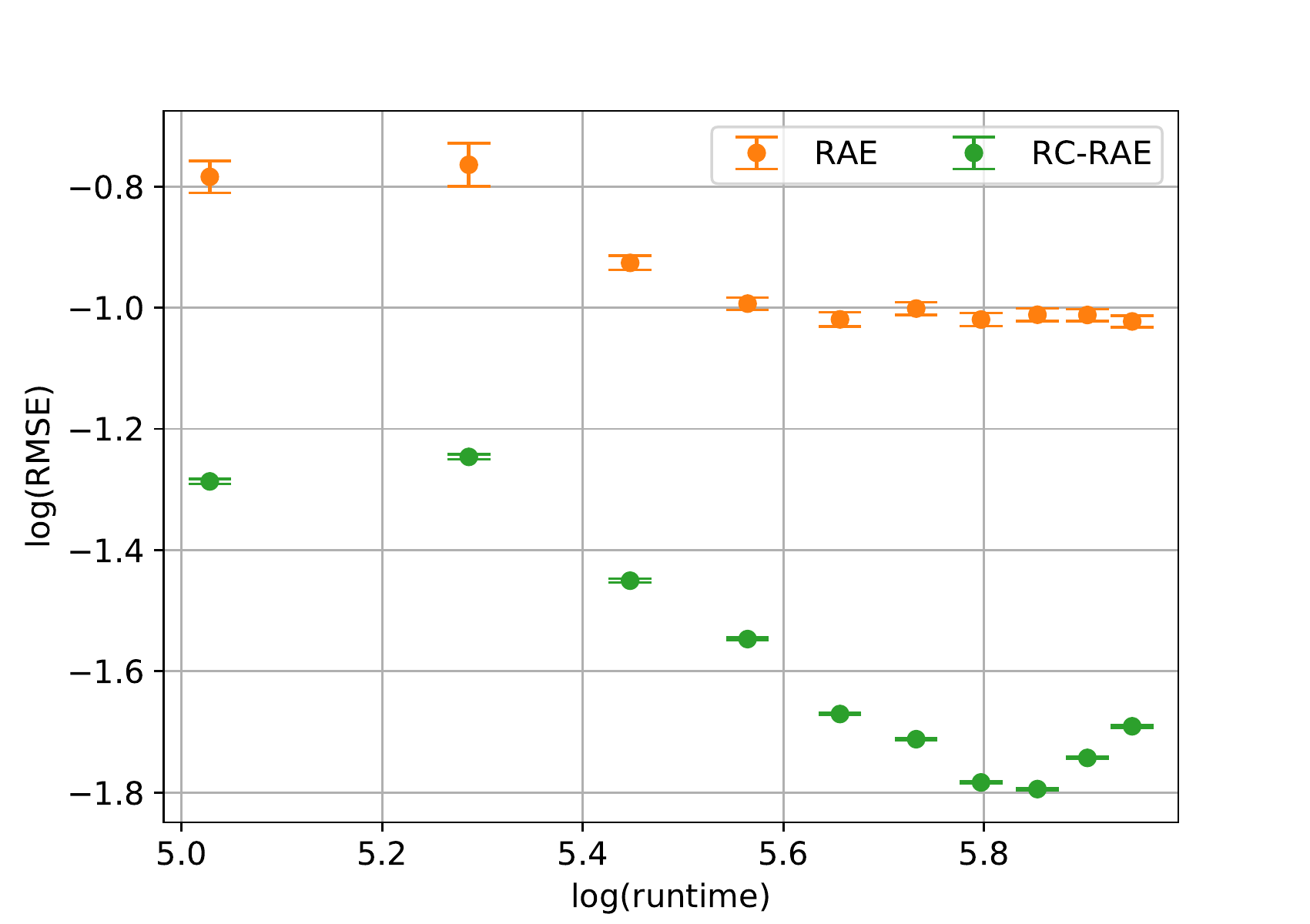}}
    \caption{Enhanced sampling properties of RAE and RC-assisted RAE (RC-RAE) as observed by simulating enhanced sampling circuits in the presence of both incoherent and coherent noise.
     Each point and its error bar represent the RMSE of 50 estimates of $\Pi= 0.2238$~(\cref{fig:compare_mitigate}) and the corresponding standard error, respectively.
  }
    \label{fig:compare_es}
\end{figure}

Through simulations of noisy quantum circuits, we study how RC impacts the two features of RAE, namely enhanced sampling for improving RMSE and error mitigation for improving bias of estimates.
In the presence of amplitude and phase damping noise and a low magnitude of residual $ZZ$ coupling, RC-assisted RAE achieves slightly lower variances and RMSEs as compared to RAE~(\cref{fig:compare_mitigate}(a)), although both techniques have equivalent scaling of RMSE with runtime~(\cref{fig:compare_es}(a)).
In this noise regime, both techniques also have comparable biases because our choice of $B$ can not resolve changes in biases at the third decimal point. 

By increasing the coherent noise to a moderate value, we observe that RC-assisted RAE performs significantly better than RAE. 
In this regime, the scaling of RMSE with runtime for RC-assisted RAE exceeds the scaling for RAE up to $L_\text{max}=7$~(\cref{fig:compare_es}(b)); this signifies that RC helps RAE to improve RMSE faster.
As we have fixed $M$ for increasing $L_\text{max}$, this perceived improvement in RMSE with increasing runtime is solely due to using more layers, and not due to increasing shots.
This feature is also evident from \cref{fig:compare_mitigate}(b) where error bars on the low-biased estimates from our RC-assisted RAE decreases with increasing $L_\text{max}$.
Moreover, due to the inherent ability of RC to mitigate the impact of coherent errors on quantum-algorithmic performance, our algorithm is able to significantly reduce bias as compared to RAE in the presence of moderate coherent noise~(\cref{fig:compare_mitigate}(b)).

Similar to our previous experiments~\cite{katabarwa2021reducing}, we notice that the estimates oscillate with $L_\text{max}$~(\cref{fig:compare_mitigate}).
This might be due to the impact of coherent error being accumulated in some systematic manner as we include additional layers in the estimation process.
These oscillations persist in RC-assisted RAE possibly due to the fact that we do not optimize the number of RC duplicates and there might exist some residual coherent errors in the circuits. 
By allocating the same number of shots for SS and RAE, we observe that RC-assisted RAE also beats SS in the presence of low to moderate coherent error by achieving far less bias and standard deviation~(\cref{fig:compare_mitigate}).


\begin{figure}[]
    \centering
    \subfloat[Error mitigation property]{
    \includegraphics[width=.85\linewidth]{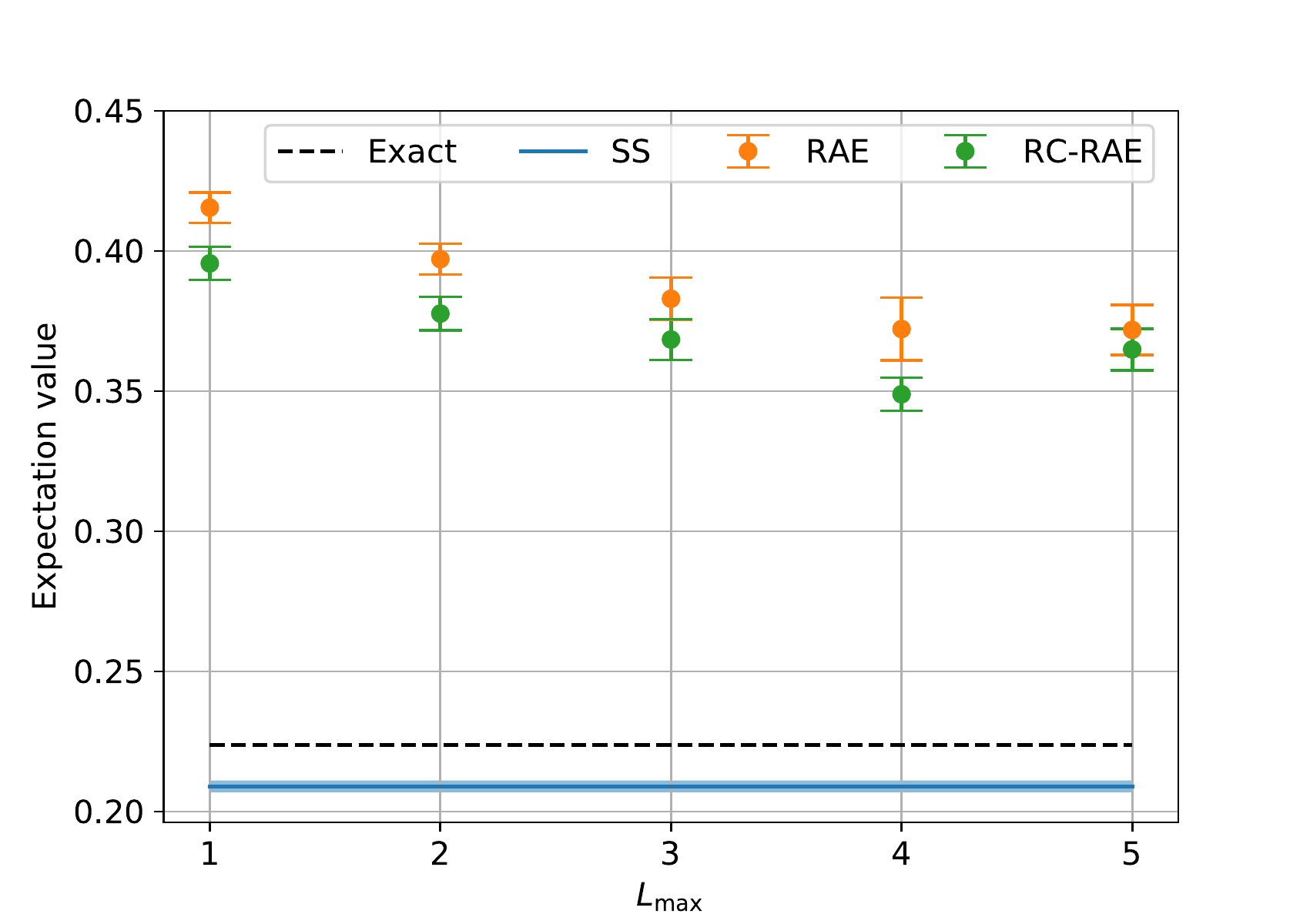}} \\
    \subfloat[Enhanced sampling property]{
    \includegraphics[width=.85\linewidth]{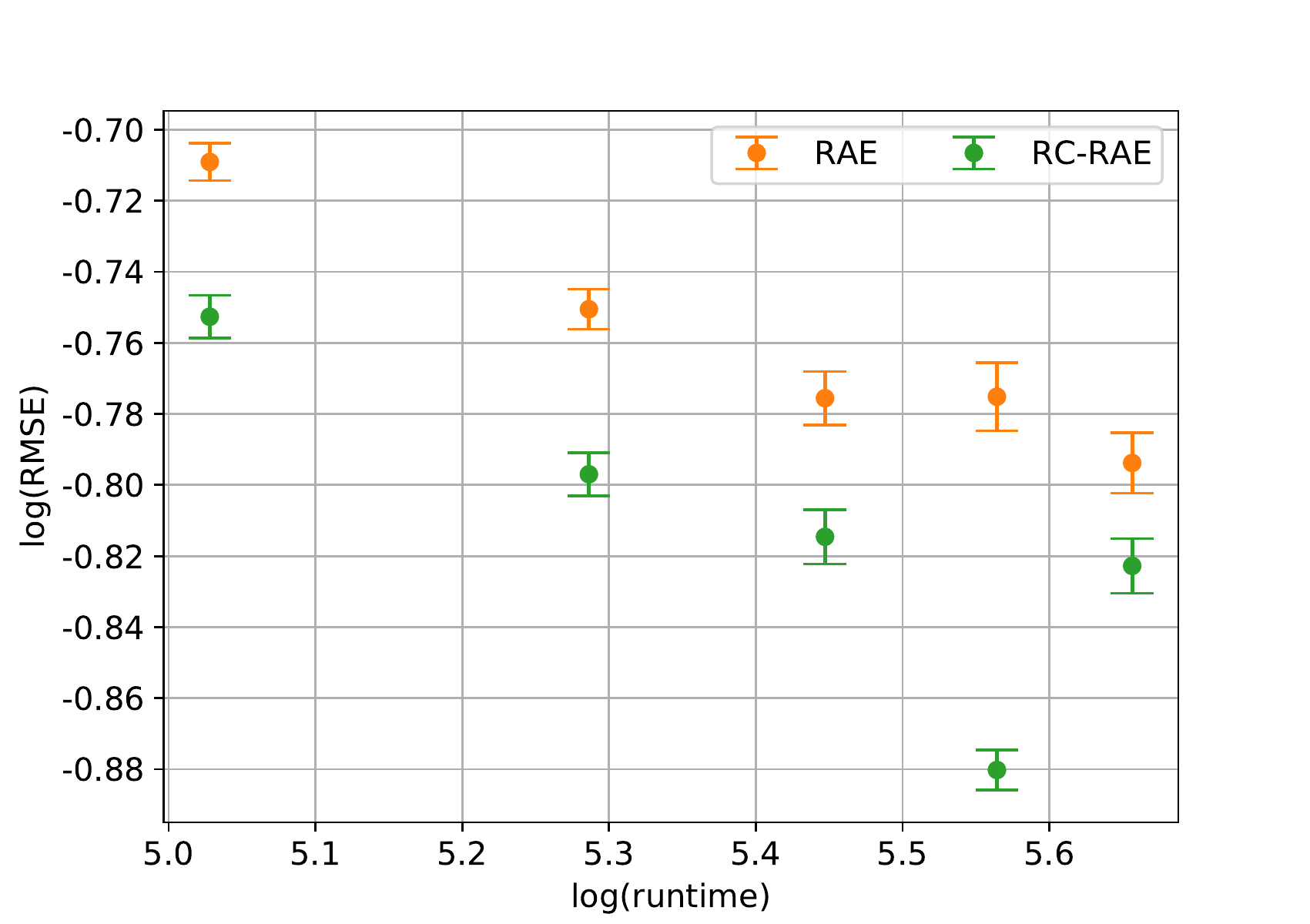}}
    \caption{Estimation of expectation values using \texttt{ibmq\_belem} for a 4-qubit Hydrogen Ansatz~(\cref{fig:4_qubit_circuit}) with $\Pi=\langle XXXX\rangle=0.2238$.
    Each point and its error bar are obtained by taking statistics over 50 independent estimates; see captions of Figs.~\ref{fig:compare_mitigate} and~\ref{fig:compare_es}.
  }
    \label{fig:expectation-hyd}
\end{figure}
\begin{figure}[]
    \centering
    \subfloat[Error mitigation property]{
    \includegraphics[width=.85\linewidth]{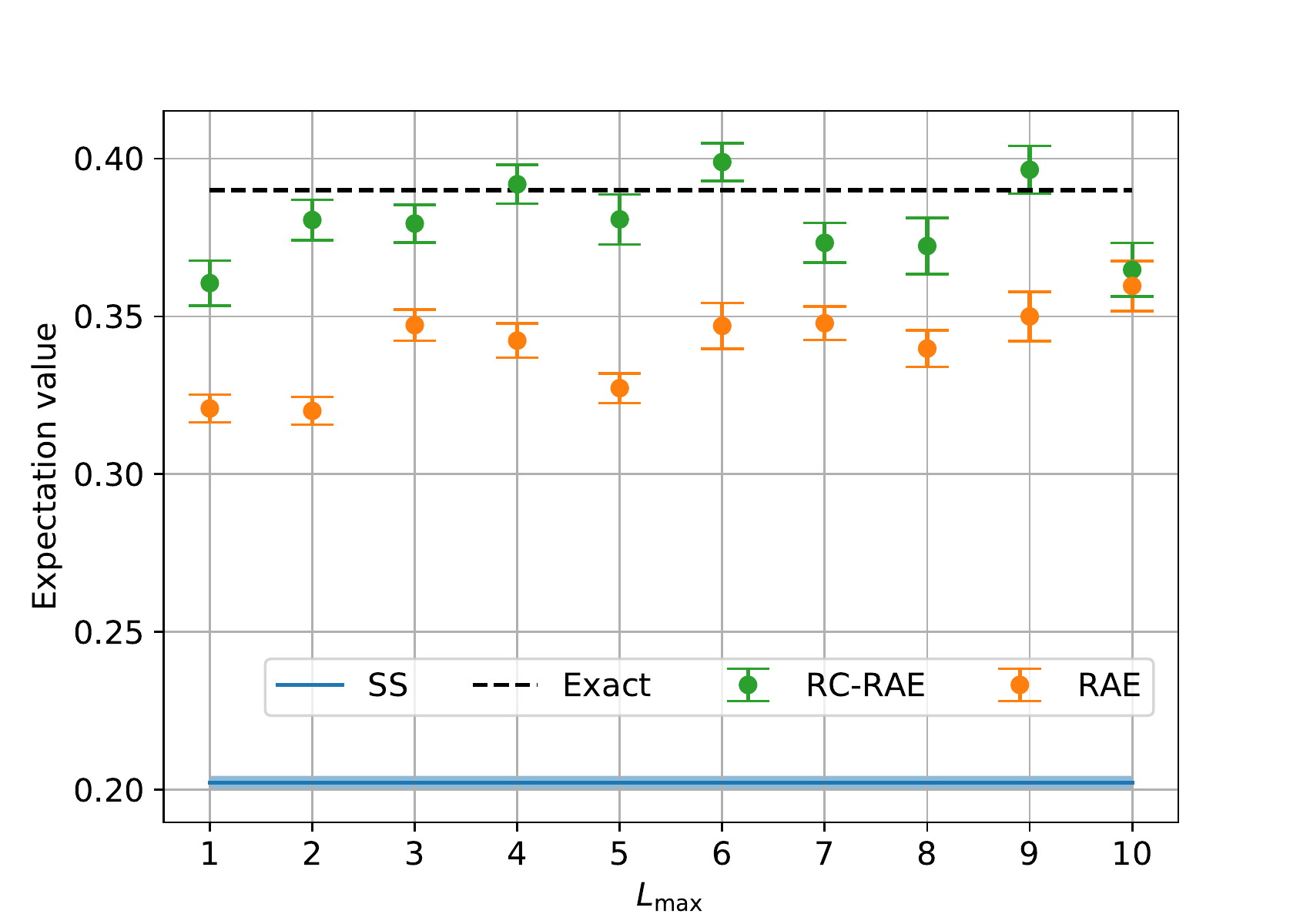}} \\
    \subfloat[Enhanced sampling property]{
    \includegraphics[width=.85\linewidth]{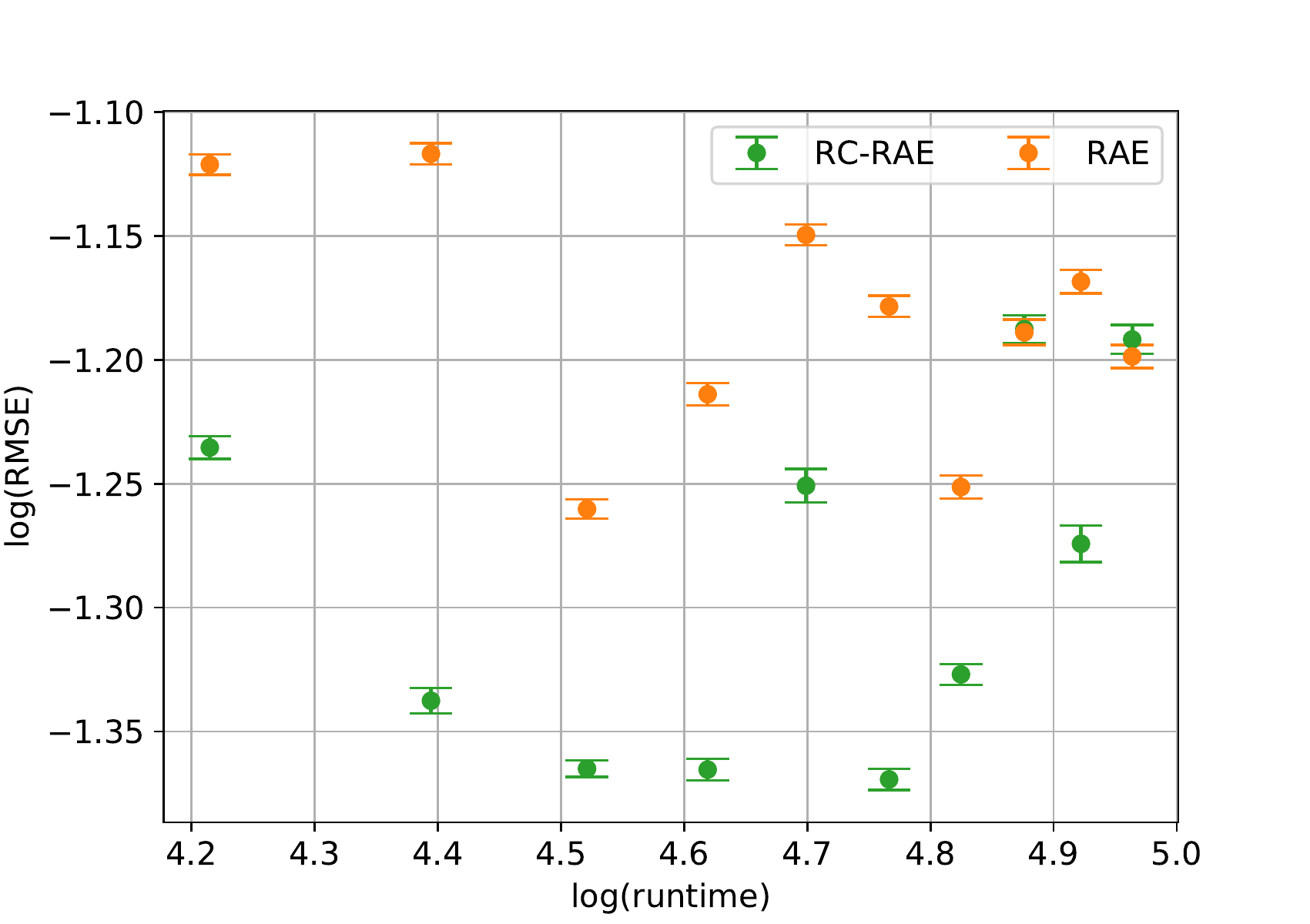}}
    \caption{Estimation of expectation value for a 2-qubit LDCA~(\cref{fig:2_qubit_circuit}), with $\Pi=\langle XX\rangle=0.39$, using \texttt{ibmq\_belem}.
Similar to~\cref{fig:expectation-hyd}, each point and its error bar are obtained by taking statistics over 50 independent estimates.
In this experiment, each estimate is obtained from $B=200000$ runs and $M=8000$ shots, with RC using $N=20$.}
    \label{fig:expectation-ldca}
\end{figure}

\subsection*{Performance on a real hardware}
Having established the importance and limitations of RC for the RAE algorithm in simulations, we now test its efficacy on real devices.
We showcase results from one publicly-available IBM device: \texttt{ibmq\_belem}.
For the 4-qubit circuits, RC-assisted RAE has only slight advantage over RAE in terms of reducing bias, but no improvement in precision is observed~(\cref{fig:expectation-hyd}(a)).
Consequently, RMSEs from both techniques scale similarly with increasing runtime, where RC-assisted RAE yields slightly less RMSEs than RAE~(\cref{fig:expectation-hyd}(b)), and this reduction of RMSEs is solely due to the corresponding decrease in biases.  
As these circuits experience huge coherent noise from various sources including residual $ZZ$ couplings and other kinds of crosstalk between qubits, the stochastic noise model achieved after Pauli twirling renders the likelihood function uninformative.
Our observation from~\cref{fig:expectation-hyd}(a) suggests a major limitation of RC for RAE: when coherent noise is very high, the effective noise model after RC doesn't assist RAE and thus we are better off opting for SS.

For the 2-qubit circuits, although we see a significant reduction in bias with RC-assisted RAE, no noticeable advantage occurs for precision of estimates~(\cref{fig:expectation-ldca}(a)). 
The impact of coherent error sources on these 2-qubit circuits is lesser than that on 4-qubit circuits, thus enabling RC to mitigate noise better.
A possible reason for getting lesser biases, both with and without RC, for 4-qubit circuits is that the readout error correction may be less effective for more qubits.
These readout errors can introduce additional bias to the estimates, which cannot be tackled by RC.
These two-qubit circuits experience a noise regime where RC-assisted RAE beats SS by reducing bias of estimates quite significantly.
Consequently, RMSEs obtained from our technique decrease up to $L_\text{max}=4$, and have lesser values than those achieved by RAE, as evident from~\cref{fig:expectation-ldca}(b).

\subsection*{Summary of results}
In practice, the performance of our RC-assisted RAE algorithm greatly depends on the impact of device noise on the enhanced sampling circuits, and consequently, this algorithm does not always deliver better performance than those from RAE or SS.
The effective stochastic noise after RC can be very detrimental when the bare circuit has a high degree of coherent error.
This is due to the fact that although RC successfully tailors the likelihood function into the desired form~\eqref{eq:noisyCheby}, the advantage of RC-assisted RAE ultimately depends on the value of $f$ for this tailored function.
A low value for $f$, corresponding to an effective stochastic noise of high strength, reduces the signal used in the estimation process and thus deteriorates the algorithmic performance.

From our experiments using simulators and real devices, we can thus infer the following:
\begin{enumerate}
    \item RC can help in improving RAE's performance if coherent errors are not high, where the high, moderate and low coherent error regimes can be algorithm dependant. 
      \item Study of optimal number of RC duplicates is needed since this might impact the bias of estimate. 
\end{enumerate}


\section{Conclusion}
\label{sec:conclusion}
The excitement of near-term quantum computing is not one for merely academics but also for practitioners in the industrial world where quantum advantage could scalably be turned into a great boon for the world at large. 
One major roadblock in achieving quantum advantage on NISQ devices is the high cost for estimating expectation values of operators with desired accuracies.
Although RAE can potentially squeeze quantum coherence from deep circuits to improve precision and bias of estimates, the presence of highly correlated noise and decoherence restricts its applicability only to small circuits.
In this work, we show that by altering complicated device noise using RC we can make RAE  more feasible on NISQ
devices.

By running RC-assisted RAE on IBM devices, both simulator and quantum processor, we propose three broad regimes for coherent noise, which is found to be the main culprit for deteriorating algorithmic performances.
In the low-noise regime, the performance of our algorithm is not conclusively better (we see disagreements in the third decimal point) than RAE's performance; this might not be surprising as there is little coherent noise to be removed and we suspect that more careful optimization of randomized compiling is needed.  But it is clear that both techniques provide significant advantage over SS.
We demonstrate these effects in simulations by modelling coherent errors as residual $ZZ$ couplings between superconducting qubits.

Upon increasing this coupling strength to a moderate magnitude, RC yields an effective stochastic noise channel, which is not very destructive.
Consequently, our RC-assisted RAE can be used to recover the sampling power of RAE and to further enhance the noise mitigating property of RAE in this regime.
As coherent error increases, RC can still help to reduce the bias but the tailored likelihood function has very little signal.
Thus, the inferencing procedure loses its sampling power and using Grover iterates does not particularly improve the scaling of the precision and RMSE, as observed by running 2-qubit circuits on today's devices. 

On the extreme end where the coherent noise gets very high, reduction in bias after RC is insignificant; thus, the noise mitigation property of RAE vanishes as well and using SS is enough. 
We find that 4-qubit circuits on NISQ devices fall in this regime.
Thus RC-assisted RAE performs differently in this high-noise regime as compared to the moderate-noise regime, where this algorithm enhances at least one of the two features of RAE.

Our work highlights the role of noise tailoring in improving performances of NISQ algorithms that rely on quantum amplitude estimation.
From our experiments on practical NISQ computers, we realize that further quantum-control techniques are essential for decreasing crosstalk errors~\cite{NLZ+22, TCK+21}, and, consequently, for extending the usefulness of our RC-assisted RAE on these devices.
Nevertheless, our technique can possibly make implementations of RAE on early fault-tolerant and fault-tolerant quantum computers perform more closely to the existing performance models~\cite{JKG+22}.
Moreover, performance of our algorithm can be improved by using optimal number of random duplicates for each enhanced sampling circuit.
A possible future direction is to combine the existing techniques for dealing with device noise (Fig.~\ref{fig:three_philosophies}) in order to recover the ability of RAE to improve precision even in particularly noisy conditions.


\section*{Acknowledgements}
We acknowledge insightful discussions on RAE and RC with Peter Johnson and Athena Caesura, respectively.
We also thank Razieh Annabestani, Jerome Gonthier and George Umbrarescu for reviewing the manuscript.
Our experiments are run using the Orquestra\textsuperscript{\texttrademark} platform of Zapata Computing.
\bibliography{paper}


\end{document}